\documentclass[prd,aps,floats,amssymb, preprintnumbers,preprint]{revtex4}
\usepackage{epsfig}

\def\new#1{#1}
\begin{document}
\preprint{\tt astro-ph/0506392}
\title{Cosmological constraints on a classical limit of quantum gravity}
\author{Damien A. Easson$^1$\footnote{easson@physics.syr.edu}, 
Frederic P. Schuller$^2$\footnote{fschuller@perimeterinstitute.ca},
Mark Trodden$^1$\footnote{trodden@physics.syr.edu} 
and Mattias N.R. Wohlfarth$^3$\footnote{mattias.wohlfarth@desy.de}}

\affiliation{$^{1)}$Department of Physics, Syracuse University, Syracuse, NY 13244-1130, USA \\
$^{2)}$Perimeter Institute for Theoretical Physics, 31 Caroline Street N, Waterloo N2L 2Y5, Canada \\
$^{3)}$Institut f\"ur Theoretische Physik, Universit\"at Hamburg, Luruper Chaussee 149, 22761 Hamburg, Germany}

\begin{abstract}
We investigate the cosmology of a recently proposed deformation of Einstein gravity, emerging 
from quantum gravity heuristics. The theory is constructed to have de Sitter space as a vacuum solution, 
and thus to be relevant to the accelerating universe. However, this solution turns out to be unstable, and the true phase 
space of cosmological solutions is significantly more complex, displaying two late-time power-law 
attractors -- one accelerating and the other dramatically decelerating. It is also shown that non-accelerating cosmologies 
sit on a separatrix between the two basins of attraction of these attractors. Hence it is impossible to pass from 
a decelerating cosmology to an accelerating one, as required in standard cosmology for consistency with 
nucleosynthesis and structure formation and compatibility with the data inferred from supernovae Ia. 
We point out that alternative models of the early universe, such as the one investigated here might provide possible 
ways to circumvent these requirements.
\end{abstract}

\maketitle

\section{Introduction}
The observed acceleration of the universe, confirmed from a cross-correlation of independent cosmological 
datasets \cite{Riess:1998cb,Perlmutter:1998np,Tonry:2003zg,Bennett:2003bz,Netterfield:2001yq,Halverson:2001yy}, 
poses a crucial question for fundamental physics. There are at least three distinct approaches from which an 
explanation of these observations might come. The simplest explanation is that cosmic acceleration is due to a 
cosmological constant. If this is the case, then our understanding of quantum fluctuations of matter fields and their 
gravitational effects will require new insights. A second possibility is that there is no cosmological term and that the 
dynamics of some new matter field, for example a scalar, lead to late-time acceleration. Such an explanation would require 
us to understand the existence of an unnaturally weakly coupled long-range field, with minute energy density at this particular 
epoch in cosmic history. A third, and less well-explored, possibility is that cosmic acceleration is due to an infrared modification of 
gravity, which yields cosmological solutions radically different from those of pure general relativity. 
While many such models 
exist \cite{Carroll:2003wy,Capozziello:2003tk,Carroll:2004de,Deffayet:2001pu,Freese:2002sq,Arkani-Hamed:2002fu,Dvali:2003rk,Arkani-Hamed:2003uy,Vollick,Easson:2004fq,Dick,Nojiri_1,Dolgov,Nojiri_2,Meng1,Chiba,Meng2,Flanagan,Woodard,Ezawa,Flanagan_2,Rajaraman,Vollick_2,Nunez,Allemandi,Lue:2003ky}, it has recently been 
shown by two of the authors that a unique theory is selected by the special geometry of Lorentzian spacetimes approximating a quantum spacetime \cite{Schuller:2004nn,Schuller:2004rn}.  

In this paper, we explore the cosmology of this distinguished deformation of Einstein gravity. The deformation takes the form
of an inverse curvature term constructed from the Riemann tensor, similar to, but distinct from the modifications discussed 
in \cite{Carroll:2004de}.
In particular we investigate in some detail the phase space structure of the resulting modified Friedmann equation.  
In ordinary Einstein gravity, if the present acceleration of the Universe is attributed to a cosmological constant, 
then any cosmology inevitably approaches the de Sitter vacuum at late times. In our case, although the theory is constructed 
to have de Sitter space as a vacuum solution, and thus to be relevant to the accelerating universe, this solution is unstable, and 
the true phase space of cosmological solutions is significantly more complex, displaying two late-time power-law attractors -- 
one accelerating and the other dramatically decelerating. However, it appears that the theory cannot explain the accelerating universe. 
This is because non-accelerating cosmologies sit on a separatrix between the basins of attraction of the two attractors. 
It is therefore impossible to pass from the decelerating cosmology required, in standard cosmology, by nucleosynthesis 
and structure formation to the accelerating one inferred from supernovae Ia. The supernovae data provides evidence for the jerk at
greater than $99\%$ confidence level \cite{Carroll:2001bv}. The phase space is divided into 
ever accelerating and ever decelerating universes; cross-overs cannot occur. 
However, for alternative cosmologies, such as the one under investigation here, the crossover 
requirement is less clear and there might be some loopholes which we will discuss.      

The structure of this paper is as follows. In the next section we describe inverse Riemann gravity and then derive the cosmological 
equations of motion necessary for the remainder of the paper. In section~\ref{phasespace} we provide an analytic phase space 
description of the most important features of the solutions. We show that the de Sitter solution is unstable, identify late-time 
power-law attractors and, most importantly, demonstrate the existence of a critical zero-acceleration separatrix. 
In section~\ref{numerics} we go on to provide complete numerical solutions to the equations of motion, mapping out 
the phase space for both the vacuum and non-vacuum cases, before concluding with a discussion in section~\ref{conclusions}. 

\section{Inverse Riemann cosmology}

\subsection{Inverse Riemann gravity}
Quantum gravity heuristics suggest that Lorentzian manifolds approximating a quantum spacetime possess both infrared and ultraviolet sectional curvature 
bounds \cite{Schuller:2004nn,Schuller:2004rn}. Such bounds are entirely equivalent to bounds on the eigenvalues of the Riemann tensor, considered as an endomorphism on the space of 2-forms. If one attempts to write down an action such that all solutions respect these bounds, one quickly discovers that there is a large class of such theories (as many as there are holomorphic functions on an annulus). The ambiguity arises because of our ignorance of the exact quantum spacetime structure in this approach. 

However, because we are interested in cosmology, it is appropriate to take the classical limit, corresponding to the removal of the 
curvature bounds. If one further requires that, in this limit, de Sitter spacetime is a solution, as one might 
hope occurs in order to explain the accelerating universe, then the resulting classical theory is unique 
\footnote{To be more precise, uniqueness of the classical limit holds within a (more tractable) subclass of all 
possible theories respecting the sectional curvature bounds, for which the Laurent series in the Lagrangian is already 
determined by a set of Taylor coefficients \cite{Schuller:2004nn}.}. Surprisingly, and interestingly, 
it is not the Einstein-Hilbert action with a positive cosmological constant. In this paper, we take the resulting classical 
action as our starting point, and proceed to investigate its broad cosmological characteristics.  

The action is a new type of infrared modification of Einstein-Hilbert gravity. Including a matter action $S_m$, 
the total action in $d$ spacetime dimensions reads
\begin{equation}\label{action}
S=\int  d^{d}\!x \, \sqrt{-g}\, \textrm{ Tr}\left(R-\frac{d-2}{d+2}\zeta^2 R^{-1}\right)+S_m\,, 
\end{equation}
where $R^A{}_B$ is the endomorphism on the space of antisymmetric two-tensors defined by the Riemann 
tensor $R^{[ab]}{}_{[cd]}$. The trace should not double-count, and hence, is defined 
by ${\rm Tr}f(R)\equiv {f(R)^A}_A = {f(R)^{[ab]}}_{[ab]}/2$. We have chosen units such that the reduced 
Planck mass is $M_P=(8\pi G)^{-1/2}=1$. The vacuum solutions are de Sitter and anti de Sitter space of constant curvature, proportional to $\pm \zeta$ respectively.

The action (\ref{action}) is similar to that introduced in \cite{Carroll:2003wy}. However, note that the current action contains both the Ricci scalar and a specific function of the full Riemann tensor, rather than just a function of the Ricci scalar. In particular, it is important to note that ${\rm Tr}(R^{-1})$ is generally not equal to the inverse of the Ricci scalar.  As noted in \cite{Schuller:2004nn}, the dimensionality of the space of metrics is different from that of the space of Riemann tensors, which means that it is not possible to perform a transformation of the metric, such as a conformal transformation, which reduces the action to that of Einstein gravity, with the extra degrees of freedom represented by additional matter fields. This means that direct comparison with solar system tests is a more complicated calculational task, which we will not attempt in this paper. It thus remains to be seen whether this action is compatible with gravitational dynamics at the solar system and galactic scales. 

The general equation of motion resulting from (\ref{action}) is
\begin{equation}\label{eqm}
R^{ij} - \frac{1}{2}g^{ij}R+\frac{d-2}{d+2}\zeta^2\left[(R^{-1})^{(i|b|j)}{}_b+\frac{1}{2}g^{ij}(R^{-1})^{ab}{}_{ab}-\nabla_b\nabla_c(R^{-2})^{c(ij)b}\right]=T^{ij}\,,
\end{equation}
where $T^{ij}$ is the matter energy-momentum tensor derived from $S_m$, and the sign convention $R^a{}_{bcd}=\partial_c\Gamma^a{}_{bd}+\Gamma^a{}_{ec}\Gamma^e{}_{bd}-(c\leftrightarrow d)$ has been used.

\subsection{Cosmological equations of motion}
We impose the cosmological Friedmann-Lema\^itre-Robertson-Walker ansatz 
\begin{equation}
\label{FLRW}
ds^2=-dt^2 +a(t)^2 d\Sigma_k^2
\end{equation}
for a homogenous and isotropic spacetime, where $a(t)$ is the scale factor and 
\begin{equation}
d\Sigma_k^2 = \bar g_{\alpha\beta}dx^{\alpha}dx^{\beta} = \frac{dr^2}{1-kr^2}+r^2\left(d\theta^2+\sin^2\theta\,d\phi^2\right)
\end{equation}
is the metric on a $d-1$ dimensional space of normalized constant curvature $k=0,\pm 1$ with Riemann tensor $\bar R_{\alpha\beta\gamma\delta}=k (\bar g_{\alpha\gamma}\bar g_{\beta\delta}-\bar g_{\alpha\delta}\bar g_{\beta\gamma})$. With this ansatz, $R^A{}_B$ is a diagonal matrix, with indices taking values in $\{[0\beta],[\alpha\beta]\}$. The corresponding $d(d-1)/2$ eigenvalues are given by
\begin{eqnarray}
R^{[0\beta]}{}_{[0\delta]} & = & \frac{\ddot a}{a}\,\delta^\beta_\delta\,,\\
R^{[\alpha\beta]}{}_{[\gamma\delta]} & = & \frac{(k+\dot a^2)}{a^2} \left(\delta^\alpha_\gamma\delta^\beta_\delta-\delta^\alpha_\delta\delta^\beta_\gamma\right),
\end{eqnarray}
the overdot denoting differentiation with respect to cosmic time $t$. 
Thus, the inverse curvature correction to the Einstein-Hilbert action appearing in (\ref{action}) is given by
\begin{equation}\label{tracer}
{\rm Tr} \, (R^{-1}) = (d-1)\left[\left(\frac{d-2}{2}\right) \frac{a^{2}}{k+ \dot a^{2}} + \frac{a}{\ddot a} \right]
\,.
\end{equation}
From the form of this correction we may already anticipate a possible difficulty in passing from a decelerating phase
to an accelerating one, since (\ref{tracer}) typically blows up when $\ddot a =0$.

We consider perfect fluid matter with the energy-momentum tensor
\begin{equation}
\label{perfectfluid}
T^{ij} = (\rho + p)u^{i} u^{j} + p g^{ij}\ ,
\end{equation} 
where $u^{i}$ characterizes the fluid's four-velocity in its rest frame, $\rho$ denotes its energy density, and $p$ its pressure. The analogue of the Friedmann equation is derived from the time-time component of the equations of motion which, after some algebra, becomes
\begin{eqnarray}
\label{friedmann}
\frac{2\zeta^2}{d+2}\left[-\frac{2a}{\ddot a}+d\left(\frac{\dot a}{\ddot a}\right)^2-\frac{2a\dot a\dddot a}{\ddot a^3} \right. &-& \left.\frac{(d-2)a^2(k+3\dot a^2)}{2(k+\dot a^2)^2}\right]\nonumber\\
&+& \frac{k+\dot a^2}{a^2} \quad = \quad \frac {2\rho}{(d-1)(d-2)}\,. 
\end{eqnarray}
Note that this result, from the full equations of motion, agrees with an effective
action calculation performed with the line element (\ref{FLRW}) additionally
including a lapse function $N(t)$ which is set to unity after variation.
Setting $\zeta\equiv  0$, one obtains the usual Friedmann equation of the standard cosmology based on the Einstein-Hilbert action. For later convenience, we adopt the usual definition of the Hubble parameter $H(t)\equiv {\dot a}(t)/a(t)$, which allows us to rewrite the above equation in the form
\begin{eqnarray}
\label{friedmannH}
\frac{2\zeta^2}{d+2}\left[-\frac{2}{\dot H+H^2}+d\left(\frac{H}{\dot H+H^2}\right)^2 \right. &-& \left. \frac{2H(\ddot H+3\dot H H+H^3)}{(\dot H+H^2)^3}-\frac{(d-2)a^2(k+3a^2H^2)}{2(k+a^2H^2)^2}\right]\nonumber\\
&+&\frac{k+a^2H^2}{a^2} \quad = \quad \frac {2\rho}{(d-1)(d-2)} \, . 
\end{eqnarray}
This equation and the continuity equation
\begin{equation}
\dot\rho+(d-1)H(\rho+p)=0\,,
\end{equation}
which follows directly from the conservation of energy-momentum $\nabla_a T^{ab}=0$, are equivalent to the full set of gravitational field equations.

\section{Phase Space Description of Solutions}
\label{phasespace}
The main goal of this work is to understand the expansion history of the universe in the theory introduced in the previous section. Because the resulting system of equations is of order greater than two, it is most convenient to understand the generic properties by considering its phase space evolution.

As is typical in cosmology, we close the system of equations by specifying a matter equation of state $p=p(\rho)$. Then, without further assumptions, the cosmological phase space of the resulting system is given by $(\rho,a,\dot a,\ddot a)$. The diffeomorphism invariance of the field equations (in particular, the time reparametrization invariance) bundles the trajectories in this phase space into classes of physically indistinguishable solutions. We may pick one member of each class by requiring that for a fixed time $t_0$ the scale factor should satisfy $a(t_0)=1$. Note that this does not change the dimension of the phase space; it merely implies that, in finding solutions, one does not need to scan over different initial conditions for the scale factor $a$.

Note that upon a global conformal rescaling $g_{ij} \mapsto \xi^{-1} g_{ij}$, the Riemann endomorphism scales with $\xi$ and the metric density with $\xi^{-d/2}$. This allows to factor out $\xi^{(2-d)/2}$ while simultaneously rescaling the deformation parameter $\zeta \rightarrow \zeta/\xi$. For a matter action that also scales with $\xi^{(2-d)/2}$, such as a massless scalar or simply the vacuum, solutions of theories with different non-zero $\zeta$ are therefore related by global conformal transformations. This means that the qualitative nature of the phase space will be the same for all values of $\zeta$.

We now survey the salient features of the phase space.

\subsection{De Sitter vacuum instability}
The equations of motion (\ref{eqm}) have precisely two solutions of constant curvature, which are the de Sitter and anti-de Sitter spacetimes with Hubble parameter (in the flat slicing) $\pm \zeta$. Note that this statement is independent of the spacetime dimension $d$.

For convenience we define $J\equiv \dot H$ and solve equation~(\ref{friedmannH}) with $k=0$ for $\ddot H$. This yields the system of equations
\begin{eqnarray}
\dot\rho & = & -(d-1)H\rho\,,\\
\dot H & = & J\,,\\
\dot J & = & \frac{(J+H^2)^3}{2H}\left\{\frac{d+2}{2\zeta^2}\left[H^2-\frac{2\rho}{(d-1)(d-2)}\right]-\frac{2}{J+H^2}\right.\nonumber\\
& & \left. +\,\frac{dH^2}{(J+H^2)^2}-\frac{2H(3HJ+H^3)}{(J+H^2)^3}-\frac{3(d-2)}{2H^2}\right\}.
\end{eqnarray}
This system possesses two fixed points. The first is at $(\rho,H,J)=(0,\sqrt{\zeta},0)$, and coincides precisely with the de Sitter vacuum solution $a(t)=\exp{\sqrt{\zeta}t}$. The second fixed point is $(\rho,H,J)=(\rho_0,0,0)$ for constant $\rho_0$. However, this fixed point satisfies $J+H^2=0$, which means that it occurs at a singularity of equation~(\ref{friedmannH}). Therefore this is not an admissible solution.

To analyze the stability properties of the de Sitter fixed point under cosmological evolution, we linearize the above equations setting $(\rho,H,J)=(x,\sqrt{\zeta}+y,z)$ with small $X^T=(x,y,z)$. As we shall see, the de Sitter solution is a saddle point in phase space, and hence unstable.

The linearized system can be written in the form $\dot X=MX$ with the matrix $M$ given by 
\begin{equation}
M=\left(
\begin{array}{ccc}
-(d-1)\sqrt{\zeta} & 0 & 0\\
0 & 0 & 1\\
-\frac{(d+2)\sqrt{\zeta}}{2(d-1)(d-2)} & (d+2)\zeta & -(d-1)\sqrt{\zeta}
\end{array}
\right).
\end{equation}
The eigenvalues of this matrix determine the type of fixed point; they are
\begin{eqnarray}
\lambda_1 & = & -(d-1)\sqrt{\zeta}\,,\\
\lambda_\pm & = & \frac{1}{2}\sqrt{\zeta}\left(-d+1\pm\sqrt{d^2+2d+9}\right).
\end{eqnarray}
Clearly, $\lambda_1<0$, implying the stability of the de Sitter solution against perturbations of the energy density $\rho$ away from the vacuum value. However, $\lambda_+>0$ and $\lambda_-<0$, implying that this solution is a saddle point, having one unstable direction in phase space. This behaviour can be seen clearly in the numerical phase space plots presented below. 

\subsection{Power law attractors}
The simplest form of accelerating expansion in the final stage of the cosmological evolution, namely a stable de Sitter vacuum, is ruled out, as we have seen. However, there exist power law attractors.

Substituting the ansatz $a\sim t^\gamma$, with constant $\gamma$, into the modified Friedmann equation (\ref{friedmann}), one finds late-time solutions ($t\rightarrow \infty$). For $k=0$, in $d>2$, we have
\begin{equation}
\gamma_\pm=\frac{1}{d+2}\left(3d\pm\sqrt{6(d^2+2)}\right).
\end{equation}
In $d=4$, these exponents are
\begin{equation}
\gamma_\pm^{(d=4)}=2\pm\sqrt{3}
\,,
\end{equation}
and correspond to effective equation of state parameters 
\begin{equation}
w_{\rm eff}^{\pm}=\frac{1\mp 2\sqrt{3}}{3}\,.
\end{equation} 
We see that $\gamma_+>1$, which describes an accelerating solution, and $0<\gamma_-<1$, which describes a decelerating solution. For $k\neq 0$, there is only the accelerating late-time solution $\gamma_+$, since the curvature term ultimately dominates, although, if inflation vastly flattens space at early times, then the decelerating behavior during the rest of cosmic history  may include an intermediate regime approximated by the decelerating solution before curvature domination in the far future. A simple linearized analysis about these solutions shows that they are both attractors, with perturbations from them decaying away at late times.

The saddle point instability of de Sitter is directly connected to the existence of the accelerating power law attractor. 
As we will see, a large class of cosmological phase space trajectories will come close to the de Sitter vacuum for a time, but ultimately 
approach power law behavior.

\subsection{The acceleration separatrix}\label{sec_separatrix}
Perhaps the most cosmologically relevant feature of the modified Friedmann equation (\ref{friedmann}) is the generic singularity at ${\ddot a}=0$. This is, in fact, the location of a phase space separatrix. To see this, note that near ${\ddot a}=0$, equation (\ref{friedmann}) takes the asymptotic form
\begin{equation}
\frac{d}{dt}{\ddot a}(t) \sim 2H(t) {\ddot a}(t) \ .
\end{equation}
Since the Hubble parameter is always positive, it is clear that already accelerating solutions (with ${\ddot a}$ small and positive) are driven to accelerate even more, and already decelerating ones (with ${\ddot a}$ small and negative) are driven to further deceleration. Thus, phase-space trajectories can never cross over from deceleration to acceleration. 

This seems to create an almost insurmountable obstacle to the construction of a realistic cosmology in which the late-time acceleration of the universe is explained by the modified gravity nature of this theory. In standard cosmology, successful nucleosynthesis and structure formation require a decelerating cosmology at early times and hence a transition from deceleration to acceleration is required by observations. 

It is worth mentioning, however, that much of the intuition we have from standard cosmology is based on the  implicit assumption that the Einstein equations (via the unmodified Friedmann equation) control the background dynamics. 
If instead, a modified Friedmann equation is used, many calculations beyond the scope of this 
paper must be redone in order to conclusively rule out a modified gravity theory. 
These include structure formation simulations and the determination of cosmological parameters, 
such as the total matter density, from a fit of the modified model to the cosmic microwave background. 
One clear example in which the standard wisdom proves inapplicable is the following: successful nucleosynthesis is 
possible even in an accelerating background if there is an asymmetry in the neutrino and antineutrino 
abundances. This is shown in~\cite{Carroll:2001bv}, without assuming a specific background dynamics.

\section{Numerical solutions}
\label{numerics}
\begin{figure}[ht]
\centering
\includegraphics[width=4.5in]{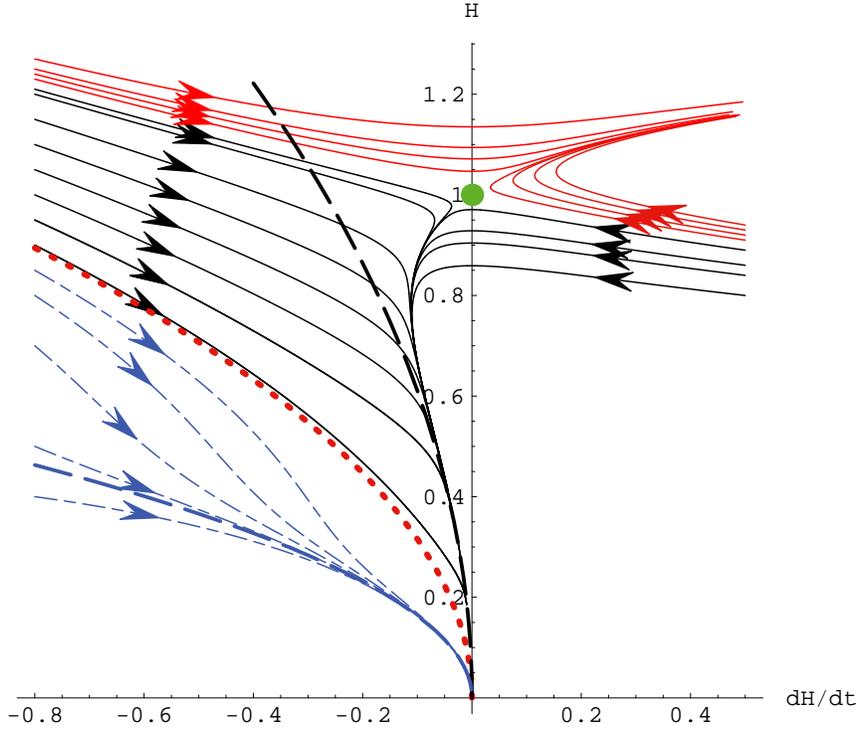}
\caption{Vacuum solutions for $\zeta=H_0=1$. The de Sitter vacuum sits on the $\dot H=0$ axis at this value (large dot). 
The two dashed curves mark the power law attractors. The accelerator is the upper of the two curves and
the decelerator is the lower one. The separatrix is indicated by the dotted line.}
\label{vacuumfigure} 
\end{figure}

While the asymptotic behavior of the theory can easily be analyzed analytically, we must solve the full equations numerically. 
To get a feeling for how solutions look, 
we restrict our attention to the case of $d=4$, flat $k=0$ cosmologies which have the most interesting phase space features. 
We first plot solutions to the vacuum equations of motion, setting $\rho=p=0$. Figure~\ref{vacuumfigure} 
shows solutions in the $({\dot H}, H)$ phase space. In the phase space plot we have scaled $\zeta$ so that the unstable de Sitter
solution is located at the point $(0, 1)$. The two late-time power law attractors and the separatrix corresponding to $\ddot a =0$  
are clearly visible in the upper left quadrant of the phase space. Solutions with power-law behavior ($a \sim t^{\gamma}$), follow the
curves $H = \sqrt{-\gamma \dot H}$ in the phase space. The upper right quadrant of the phase space is included for
completeness, however, these solutions have $\dot H > 0$ and are physically irrelevant for the purposes of our paper.

\begin{figure}[ht]
\centering
\includegraphics[width=4.5in]{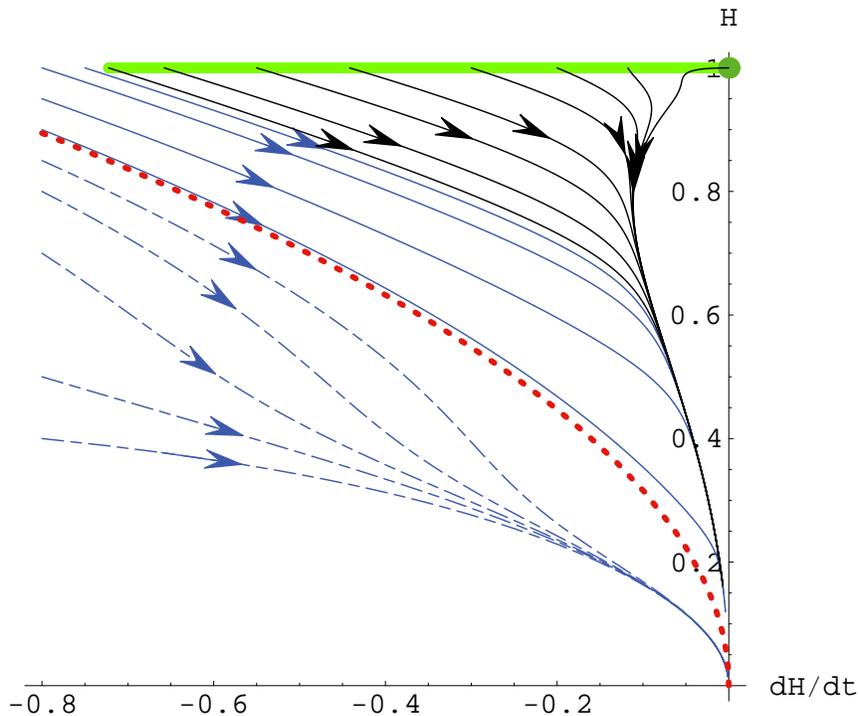}
\caption{Solutions to equation~(\ref{friedmann}), with $\Omega_m^{(0)} =0.28$, for a variety of different initial conditions. 
The thick solid line denotes initial conditions that are consistent with constraints
on the evolution of the scale factor from observations of type Ia supernovae. 
These constraints exclude the other solutions at the $95\%$ confidence level.} 
\label{matterfigure}
\end{figure}

Since we are interested in the late universe, we now specialize to the case in which the dominant matter component is pure dust, 
for which the pressure vanishes $p=0$. We set $\Omega_m^{(0)} =0.28$, $a_0=1$ and $H(t=t_0)=H_0$, where a subscript $0$ 
denotes the value of a quantity today. Figure~\ref{matterfigure} 
shows these solutions in the same $({\dot H}, H)$ phase space.

We focus on solutions with initial conditions $\rho(t_0)$ and $\dot a(t_0)$, obtained from observations of the current values 
of the Hubble parameter and the density parameter $\Omega_m\equiv H^{-2}\rho/(d-1)$, consistent with the present state of the universe
within the context of Einstein gravity. 
We therefore set $\rho(t_0)=3 H_0^2 \Omega_m,\,a(t_0)=1$ and $\dot a(t_0)=H_0$.
It is then sufficient to scan over initial conditions for the remaining phase space variable $\ddot a$.
Since we are in four dimensions, we only exhibit projections to three-dimensional hypersurfaces of 
constant $a\equiv 1$. For the present values of the matter density and the Hubble parameter we 
choose $\Omega_m=0.28$ and $H_0=1$. To find initial conditions for $\ddot a$ (or equivalently $\dot H$) compatible
with the recent supernovae data, we assume Einstein gravity and a flat universe filled with dust ($\Omega_{m}$) and a ``dark energy"
component with effective instantaneous equation of state parameter $w_{0}$.  From the Einstein equations we find 
\begin{equation}\label{ichdot}
\dot H = \frac{3}{2}H^{2} \left( 1 + \frac{w_{0}}{1+\alpha} \right)
\,,
\end{equation}
where 
\begin{equation}
\alpha = \frac{\Omega_{m}}{1-\Omega_{m}}
\,.
\end{equation}
The initial conditions for $\dot H$, compatible with supernovae data are located in the green line in Figure 2, and are obtained 
using (\ref{ichdot}) together with the observational data $-1.48 < w_{0}< -0.72$ at the $95 \%$ confidence 
level \cite{astro-ph/0205096,astro-ph/0211522,Tonry:2003zg,astro-ph/0407259}.

\section{Conclusions}
\label{conclusions}
The discovery of cosmic acceleration seems to imply that there is something important that we do not understand about particle physics, 
gravity or the interaction between them. While there have been many attempts to understand this phenomenon at a phenomenological level, it would 
be extremely exciting to derive the late-time evolution of the universe from a fundamental theory. 

In this paper we have investigated the cosmological evolution of the 
classical modified gravity theory given by the action (\ref{action}). 
This theory arises in the following way: considerations of quantum gravity heuristics suggest that 
Lorentzian manifolds approximating a quantum spacetime possess sectional curvature 
bounds~\cite{Schuller:2004nn,Schuller:2004rn}. A large class of dynamics whose solutions respect these 
curvature bounds exists. For a certain subclass, the classical limit corresponding to the removal of the 
curvature bounds, leads to an interesting result: one obtains a unique theory containing de Sitter 
space as a vacuum solution. The action for this theory is not the Einstein-Hilbert action with a positive 
cosmological constant, but rather contains a correction in terms of the inverse eigenvalues of the Riemann 
tensor.

Using the resulting action as our starting point, we have investigated its broad cosmological characteristics. As is typical of modified gravity 
theories with inverse curvature terms~\cite{Carroll:2004de,Easson:2004fq}, the equations of motion contain terms that are singular for certain 
values of the metric components. In the cosmological setting this translates into certain combinations of time derivatives of the scale factor. 
In many models~\cite{Carroll:2004de,Easson:2004fq} these singularities lie in regions of the parameter space that are never reached by 
trajectories corresponding to realistic cosmological evolution. However, in the specific theory \new{} defined by the action~(\ref{action}), 
the relevant singularity of the equations of motion occurs precisely when ${\ddot a}(t)$=0. 

If one interprets the available cosmological data sets under the assumption that Einstein gravity controls the 
dynamics of the background, there is good evidence that the universe underwent a decelerating phase 
(needed for successful nucleosynthesis and structure formation), and now a late-time accelerating phase 
(to fit, among other data, the observations of type Ia supernovae). Thus standard cosmology requires that the 
universe pass from deceleration to acceleration at some point in cosmic history. This is not allowed by the 
singularity structure of the theory investigated here.

Assuming that, even with the modified Friedmann equation (\ref{friedmann}), 
nucleosynthesis and structure formation can only take place during a decelerating phase, as in the
standard Big-Bang model, the exclusion of cross-overs from decelerated to accelerated expansion 
presents an insurmountable obstacle to a viable cosmology. It is worth commenting, however, that in order to 
reliably confront alternative cosmologies with observations, one must either use data without assuming 
specific background dynamics, as is done e.g. in \cite{Carroll:2001bv,Visser:2004bf}, or redo 
standard calculations (such as extracting the cosmological parameters from data on the basis of the modified dynamics). 
This of course presents a formidable task and is beyond the scope of this paper.

\acknowledgments
The authors thank Joshua Goldberg, Stefan Hofmann and Wilfried Buchm\"uller for helpful comments. 
The work of MT and DE is supported in part by the National Science Foundation under grant PHY-0354990, by funds provided by 
Syracuse University and by Research Corporation. 


\end{document}